\documentclass{article}

\usepackage[preprint,nonatbib]{neurips_2022}

\usepackage[utf8]{inputenc} 
\usepackage[T1]{fontenc}    
\usepackage{hyperref}       
\usepackage{url}            
\usepackage{booktabs}       
\usepackage{amsfonts}       
\usepackage{nicefrac}       
\usepackage{microtype}      
\usepackage{xcolor}         
\usepackage{graphicx}
\usepackage{orcidlink}
\usepackage{siunitx}

\newcommand{\vol}{V_i}

\newcommand{\rot}{R_i}
\newcommand{\trans}{\mathbf{t}_{i}}

\newcommand{\projgt}{Y_i}
\newcommand{\enc}{\psi}
\newcommand{\psf}{P_i}
\newcommand{\ctf}{C_i}
\newcommand{\shift}{T}
\newcommand{\fshift}{\mathcal{T}}

\newcommand{\projpred}{\hat{Y}_i}
\newcommand{\fprojpred}{\hat{\mathcal{Y}}_i}
\newcommand{\surf}{I_i}
\newcommand{\fsurf}{\mathcal{I}_i}

\newcommand{\noise}{Z_i}
\newcommand{\fnoise}{\mathcal{Z}_i}

\newcommand{\xref}{X^{(0)}}
\newcommand{\Ud}{\tilde{U}_d}

\newcommand{\conf}{z_i}
\newcommand{\confspace}{\mathcal{V}}

\newcommand{\N}{N}

\newcommand{\zdim}{d}

\newcommand{\res}{D}
\newcommand{\feature}{y_i}

\title{Heterogeneous reconstruction of deformable atomic models in Cryo-EM}

\author{
Youssef Nashed\orcidlink{0000-0001-6146-3939}\\
SLAC National Accelerator Laboratory\\
\And
Ariana Peck\orcidlink{0000-0002-5940-3897}\\
SLAC National Accelerator Laboratory\\
\And
Julien Martel\orcidlink{0000-0002-4928-5487}\\
Stanford University\\
\And
Axel Levy\orcidlink{0000-0001-7890-9562}\\
Stanford University\\
\And
Bongjin Koo\orcidlink{0000-0002-3611-4988}\\
University of California Santa Barbara\\
\And
Gordon Wetzstein\orcidlink{0000-0002-9243-6885}\\
Stanford University\\
\And
Nina Miolane\orcidlink{0000-0002-1200-9024}\\
University of California Santa Barbara\\
\And
Daniel Ratner\orcidlink{0000-0002-5747-7323}\\
SLAC National Accelerator Laboratory\\
\texttt{dratner@slac.stanford.edu}
\And
Fr\'{e}d\'{e}ric Poitevin\orcidlink{0000-0002-3181-8652}\\
SLAC National Accelerator Laboratory\\
\texttt{frederic.poitevin@stanford.edu}
}

\begin{document}

\maketitle

\begin{abstract}
    Cryogenic electron microscopy (cryo-EM) provides a unique opportunity to study the structural heterogeneity of biomolecules. Being able to explain this heterogeneity with atomic models would help our understanding of their functional mechanisms but the size and ruggedness of the structural space (the space of atomic 3D cartesian coordinates) presents an immense challenge. Here, we describe a heterogeneous reconstruction method based on an atomistic representation whose deformation is reduced to a handful of collective motions through normal mode analysis. Our implementation uses an autoencoder. The encoder jointly estimates the amplitude of motion along the normal modes and the 2D shift between the center of the image and the center of the molecule . The physics-based decoder aggregates a representation of the heterogeneity readily interpretable at the atomic level. We illustrate our method on 3 synthetic datasets corresponding to different distributions along a simulated trajectory of adenylate kinase transitioning from its open to its closed structures. We show for each distribution that our approach is able to recapitulate the intermediate atomic models with atomic-level accuracy.
\end{abstract}

\section{Introduction}
\label{sec:intro}
The cellular machinery functions thanks to molecular populations constantly transitioning between metastable states --- or \textit{conformations}. 
As a result, understanding the structural heterogeneity of biomolecules is key to understanding these essential building blocks of life~\cite{alberts2017molecular,landau2013statistical} and to subsequently making progress in drug and molecular design~\cite{johansson2018structural,wigge2020rapidly}. Cryogenic electron microscopy (cryo-EM) of molecules~\cite{glaeser2021single} is the leading technique for generating large imaging datasets of biomolecules at equilibrium. Each 2D image in the dataset is unlabelled and contains a unique molecule that adopts an unknown 3D \textit{conformation} and an unknown \textit{pose} (orientation and position).

While \textit{homogeneous} reconstruction methods focus on estimating the \textit{average volume} of the studied molecule from collected images~\cite{Scheres2012RELION, Punjani2017CryoSPARC:Determination, MiolanePoitevin2020CVPR, gupta2021cryogan, Nashed2021endtoend, levy2022cryoai}, \textit{heterogeneous} methods take into account structural variability and introduce a latent variable (the \emph{conformational state}) that characterizes the \textit{volume} associated with each observation.
Given the image formation model, heterogeneous reconstruction can be formulated as an inverse problem where the volume representation, conformation and pose must be inferred from the images~\cite{Donnat2022}.
Most heterogeneous methods to date treat the variability in the dataset on purely statistical grounds, through expectation-maximization approaches~\cite{Scheres2012RELION} or inference in deep generative models \cite{Zhong2019ReconstructingModels}. However, emerging approaches also inject knowledge about the underlying physics of the molecule in the forward model of the solver - specifically, in the way deformations of the volume are modelled (the \emph{dynamic} model). Punjani \textit{et al.} \cite{Punjani2021} and Herreros \textit{et al.} \cite{Herreros2021} model the deformation of the voxelized volume with convective fields. Zhong \textit{et al.} \cite{zhong2021exploring} and Chen \textit{et al.} \cite{Chen2021} use an initial coarse-grained atomic model as a reference from which the pseudo-atoms can deviate with a penalty. Rosenbaum \textit{et al.} \cite{Rosenbaum2021} keep all the atoms and the molecule is segmented in rigid blocks (corresponding to the protein residues) that can also deviate from a reference with a penalty, followed by minimization of the molecule's potential energy, parameterized with an all-atom pairwise interaction force-field. While promising, this approach suffers from the very large size of the structural space (scaling with the number of residues, e.g. $10^2-10^4$) and expensive energy minimization steps at each iteration for every image. 
Performing a normal mode analysis (NMA)~\cite{levitt1985protein} of the molecule represented with an elastic network model (ENM)~\cite{tirion1996large} is a popular approach to modelling the essential dynamics of a molecule at a reduced problem dimensionality. In this representation, the size of the conformational space could in principle be reduced to a handful of normal modes while preserving the ability to model relevant functional motions of the molecule~\cite{tama2001conformational}. This approach was implemented in \cite{Jin2014}, where the pose and normal mode amplitudes are iteratively estimated for each image of the dataset. However, as the authors note, this approach suffers from the same scalability issue as expectation-maximization approaches (e.g. \cite{Scheres2012ADetermination}). To remedy this, two approaches could be taken. The first approach would consist of applying this method to a subset of the data (the training set) and using its output to train a neural network, which would then infer the latent variables for the remaining data --- this supervised approach was implemented and tested on synthetic and experimental datasets in Hamitouche and Jonic \cite{hamitouche2022deephemnma}.

Here we introduce an unsupervised alternative that uses direct gradient-based optimization to jointly estimate the translational component of the pose (the shift) and the normal mode amplitudes of the molecule. Our method uses an autoencoder pipeline~\cite{Goodfellow-et-al-2016}. The encoder associates each particle image with a shift and a normal mode amplitude. The decoder is a generative physics-based pipeline that uses the predicted shift and normal mode amplitude along with the reference atomic model to predict an image. By learning a mapping from images to shifts and normal mode amplitudes, this technique avoids the computationally expensive step that limits existing cryo-EM reconstruction methods. Our approach thus amortizes over the size of the dataset and provides a scalable approach to working with modern, large-scale cryo-EM datasets~\cite{levy2022cryoai}.

\section{Approach}
\label{sec:approach}
\subsection{Image Formation Model}
\label{subsec:ifm}

In single particle cryo-EM, probing electrons interact with the electrostatic potential created by the molecules embedded in a thin layer of vitreous ice. Each molecular volume $\vol$ can be seen as a mapping from $\mathbb{R}^3$ to $\mathbb{R}$ and is indexed by $i\in\{1,\ldots,\N\}$. 
In the sample, each molecule is in an unknown orientation $\rot\in SO(3)\subset \mathbb{R}^{3\times 3}$, and an unknown conformation $z_i \in \mathbb{R}^\zdim$. We assume that the volumes $\{V_i\}_{i=1,\ldots,\N}$ are drawn independently from a probability distribution $\mathbb{P}_V$ supported on a low-dimensional subspace (the \textit{conformational space}). More specifically, we assume there exist $\zdim\in\mathbb{N}$ and $\confspace:\mathbb{R}^3\times\mathbb{R}^\zdim\to\mathbb{R}$ such that $\mathbb{P}_V$ is supported on the conformational space $\{\confspace(.,z), z\in\mathbb{R}^\zdim\}$.

\paragraph{Elastic Network Models, Normal Mode Analysis and Coarse-Graining.} The structural space of the molecular structure is $3M$ dimensional, noting $M$ its number of constitutive atoms $\sim 10^2-10^5$. Efficient sampling of this very high-dimensional space to uncover the lower-dimensional subspace defined by a handful of collective variables (the \textit{conformational space} $\confspace$) has been an area of intense research for many decades \cite{schutte1999direct, levitt2014birth, noe2017collective}. A relatively cheap and efficient option consists in performing a harmonic approximation of the energy landscape around a chosen conformation~\cite{noguti1982collective,levitt1985protein} by representing the molecule with an elastic network model (ENM)~\cite{tirion1996large,atilgan2001anisotropy}. Noting $X = \{\mathbf{r}_j\}_{j=1,M}$ the atomic Cartesian coordinates of the molecule, the potential energy $E$ of the ENM is further approximated with a second-order Taylor approximation around a reference conformation $\xref$ and using $H$ the Hessian matrix of $E$ in $\xref$:
\begin{equation}
    E(X) = \frac{1}{2}(X-\xref)^{T} H (X-\xref)~\mbox{ with }~H_{jk} \propto (\mathbf{r}_j^{(0)} - \mathbf{r}_k^{(0)})(\mathbf{r}_j^{(0)} - \mathbf{r}_k^{(0)})^{T}.
\end{equation}
Solving the equations of motion in $E$ leads to the eigenvalue problem $HU = \Lambda U$ where each of the $3M$ eigenpairs $(\lambda_l, U_l)$ defines a \textit{normal mode} of frequency $\sqrt{\lambda_l}$ that deforms the reference conformation along $U_l = \{\mathbf{u}_j^{(l)}\}_{j=1,M}$. Pragmatically, the normal modes provide a linear basis of functions that spans the whole conformational space. Because the variance of each mode scales with the inverse of its frequency, it is customary to perform a low-rank approximation of $H$ by only considering the $d$ lowest frequency normal modes $\Ud = \{U_1, ..., U_d\}$ - a choice empirically justified as they often encode relevant functional motions~\cite{tama2001conformational}. Formally, for rank $d$, the reference atomic model can be deformed as follows:
\begin{equation}
\label{eq:lindef}
    X(z) = \xref + \Ud.z^{T}.
\end{equation}
In practice, the eigenvalue problem is challenging to solve for large $M$ as the Hessian is a $3M\times 3M$ matrix. Because the lowest-frequency modes usually exhibit strong collectivity~\cite{tama2001conformational}, $H$ is usually computed for a subset of atoms, the \textit{coarse-grained (CG) model}. Its eigenvectors can then be interpolated on the remaining set of atoms and orthonormalized to yield $\Ud$.

\paragraph{Isolated Atom Superposition Approximation, Weak-phase Approximation and Projection Assumption.} The electrostatic potential of the whole molecule can be approximated as the superposition of the electron form factors of its constitutive atoms~\cite{VULOVIC201319}. Each form factor is defined as the sum of 5 Gaussians --- their amplitude $a$ and width $b$ determined by the atomic type \cite{colliex2006electron}.
For typical cryo-EM experiments, the scattered wave can be linearized and simplifies to the integral of the rotated volume along the path of the probing electron \cite{vulovic2014use}, resulting in the following mapping from $\mathbb{R}^2$ to $\mathbb{R}$ where $x_j^{(i)}$ and $y_j^{(i)}$ are the planar coordinates of atom $j$ in volume $i$ derived from $\mathbf{r}_j^{(i)} = R_i.(\mathbf{r}_j^{(0)} + \sum_{l=1}^{d} z_l^{(i)}\mathbf{u}_j^{(l)})^T$ :

 \begin{equation}
 \label{eq:proj}
     \surf(x,y) = \int_t  V_i(x,y,t)dt = 4\pi \sum_{j=1}^{M}\sum_{k=1}^{5}\frac{a_{jk}}{b_{jk}}\ \exp{-\frac{4\pi^{2}}{b_{jk}}\big((x-x_j^{(i)})^{2} + (y-y_j^{(i)})^{2}\big)}.
\end{equation}

The interaction between the beam and the microscope's lens is modeled by the Point Spread Function (PSF) $P_i$. Imperfect centering of the molecule in the image is characterized by small translations $\trans\in\mathbb{R}^2$. Finally, taking into account signal arising from the vitreous ice in which the molecules are embedded as well as the non-idealities of the lens and the detector, each image $\projpred$ can be modelled as~\cite{VULOVIC201319,Scheres2012RELION}
\begin{eqnarray}
    \projpred = \shift_{\trans} * \psf * \surf + \noise ~~~\Leftrightarrow~~~ \fprojpred = \fshift_{\trans} \odot \ctf \odot \fsurf + \fnoise
\label{eq:ifm}
\end{eqnarray}
where $*$ is the convolution operator, $\odot$ indicates element-wise multiplication, $\shift_\mathbf{t}$ is the $\mathbf{t}$-translation kernel and $\noise$ white Gaussian noise on $\mathbb{R}^2$. In Fourier space, we note $\fprojpred$ the Fourier transform of $\projpred$, $\fsurf$ the complex Fourier transform of the ideal image $\surf$,  $\fshift_\mathbf{t}$ the $\mathbf{t}$-translation operator or phase-shift in Fourier space, $\ctf$ is the Contrast Transfer Function (CTF), Fourier transform of the PSF, and $\fnoise$ the Fourier transform of $\noise$.

\subsection{Overview of the Architecture}
\label{subsec:ovw}

\begin{figure}[t!]
  \centering
  \includegraphics[width=0.9\textwidth]{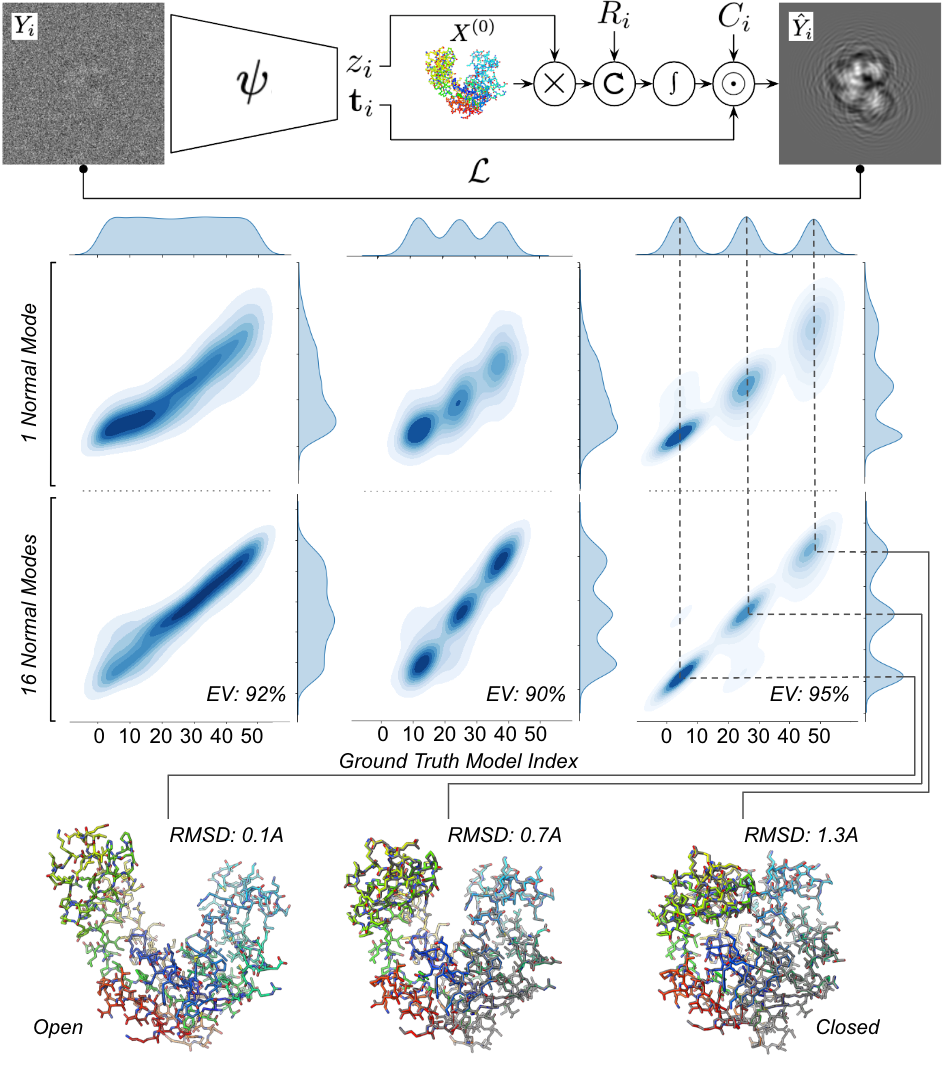}
  \caption{\textbf{(top) Autoencoder architecture}. The encoder, parameterized by $\enc$, maps an image $\projgt$ to the shift $\trans$ and conformational $\conf$ variables. The decoder is a differentiable simulator that generates noise-free images $\projpred$ using the image formation model defined in Eq.~\ref{eq:ifm} from a reference atomic model $X^{(0)}$, using the variables learned by the encoder. The input and output images are compared with the L2 loss, which is batch-minimized for $\enc$. \textbf{(bottom) Results}. GT and predicted distributions over the \textit{uniform} (left), \textit{continuous} (center) and \textit{discontinuous} (right) datasets (see Section. \ref{sec:results}) for $d=1$ (top row) and $d=16$ (bottom row). The explained variance (EV) of PC1 is indicated for $d=16$. The GT (in gray) and $d=16$ predicted atomic models (rainbow-colored) corresponding to the local maxima in the \textit{discontinuous} dataset are displayed together with their RMSD.} 
  \label{fig:architecture}
\end{figure}

Fig.~\ref{fig:architecture} summarizes the proposed encoder-decoder architecture. Images $\projgt$ are fed into an encoder, parameterized by $\enc$, that predicts a shift $\trans$ and a conformational variable $\conf$. These estimated parameters are then applied to a reference structure as in Eq.~\ref{eq:lindef}. The resulting deformed atomic model is rotated using a known pose $R_i$, and Eq.~\ref{eq:proj} is evaluated at the desired pixel positions. Based on the estimated translation $\trans$ and given CTF parameters $\ctf$, the rest of the image formation model described in Eq.~\ref{eq:ifm} is simulated with the physics-aware decoder to obtain $\projpred$, a noise-free estimation of $\projgt$. Pairs of measured and reconstructed images are compared using the L2 loss, and gradients are backpropagated throughout the differentiable model in order to optimize the encoder.

\paragraph{Discriminative model.} The encoder acts as a discriminative model by mapping images to estimates of $\trans$ and $\conf$. The encoder is structured sequentially with the following components. A Convolutional Neural Network (CNN) containing $5$ blocks, each consisting of $2$ \emph{Conv3x3} layers, followed by batch normalization and average pooling by a factor of $2$. The filter numbers for the convolutional layers in the blocks are set as [32, 64, 128, 256, 512]. The architecture of the CNN is inspired from the first layers of VGG16~\cite{vgg16}, known to perform well on visual tasks. A \emph{translation} Multi-Layer Perceptron (MLP) with $2$ hidden layers of sizes [512, 256] that outputs a feature $\trans$ of dimension $2$. A \emph{conformation} MLP with $2$ hidden layers of sizes [512, 256] that maps $\feature$ to $\conf$ of dimension $d$, which is the number of normal modes used.

\paragraph{Generative Model.} The generative model implemented in the decoder is a simulation of the image formation model (Section~\ref{subsec:ifm}). 
\textit{Pre-processing}. A reference atomic model $X^{(0)}$ is fed to ProDy~\cite{bakan2011prody} to pre-compute the Hessian matrix $H_{CG}$ of the CG model. $H_{CG}$ is diagonalized on the GPU~\cite{pytorch} and its first $d$ eigenvectors are interpolated on the remaining set of atoms with PyKeOps~\cite{pykeops} to yield $U_d$. The electron form factors~\cite{colliex2006electron} are retrieved with Gemmi~\cite{Wojdyr2022}. For each image $i$, the rotation matrix and CTF are provided either by the simulator or by an external software like RELION~\cite{Scheres2012RELION} or cryoSPARC~\cite{Punjani2017CryoSPARC:Determination}. 
\textit{Training}. The forward pass is differentiable with respect to $\conf$ and $\trans$. For each image $i$, $X^{(0)}$, $\Ud$ and $z_i$ are combined according to Eq.\ref{eq:lindef}. The resulting atomic coordinates are rotated by $\rot$ \emph{via} a matrix multiplication. A fixed grid of $\res^2$ coordinates on the x-y plane in real space is used to evaluate Eq.\ref{eq:proj}. The resulting image is Fourier transformed and element-wise multiplied by $\hat{T}_{\trans}$ and $\ctf$. A final inverse Fourier transform step is carried out.

\paragraph{Training Procedure.} 
 For all described datasets, the data was split into training and test set with a 90-10 ratio, resulting in 45,000 images in the training set and 5,000 images in the test set. Training was carried with the following parameters: we use the Adam optimizer~\cite{Kingma2015Adam:Optimization} with a learning rate of \SI{1e-3}, minibatch size of 256 images. Each training run was stopped after 200 epochs, each taking approximately 10 minutes on 4 Nvidia Tesla A100 GPUs. Inference results below are reported on the test set.

\section{Results}
\label{sec:results}
%
We generated a 50-frame trajectory between the atomic models of the open (4AKE~\cite{muller1996adenylate}) and closed (1AKE~\cite{muller1992structure}) forms of Adenylate Kinase (AK) using the \textit{morph} tool in PyMOL~\cite{PyMOL}. For each frame, 10,000 images were generated using our simulator, each image is $192\times192$ with a $0.8$\r{A} pixel size. Particle poses were randomly sampled from a uniform distribution on $SO(3)$, and CTF defocus values were randomly sampled from a Log-normal distribution with $\mu=\SI{1}{\micro\metre}$ and $\sigma=\SI{0.3}{\micro\metre}$. Gaussian noise was added to each image with a Signal-to-Noise Ratio (SNR) set at $-20$ dB. A variable number of images for each frame were then combined to generate 3 datasets, each consisting of 50,000 images, displaying different distributions (following the approach of \cite{Zhong2019ReconstructingModels}): (i) \textit{uniform} --- approximately the same number of images across frames, (ii) \textit{continuous} --- 3 overlapping Gaussians, and (iii) \textit{discontinuous} --- 3 non-overlapping Gaussians.

We tested our proposed architecture on each dataset, using AK in the open form a $\xref$. For each image $i$, $R_i$ and $\ctf$ are given from the simulation while $\trans$ and $\conf$ are estimated by the encoder $\enc$. We ran three  simulations in each case, with $d=0$ (no deformation), $d=1$ and $d=16$. For $d=1$, we compare in Fig.\ref{fig:architecture}-bottom the distribution of $\conf$ to the ground-truth (GT) distribution. For $d=16$, we first perform principal component analysis (PCA) on $\conf$ and display the first component (PC1) against GT. In every case, PC1 explained more than 90\% of the variance in $\conf$. 

While the qualitative features of the GT distribution is recaptured for $d=1$ already, the predicted distribution deteriorates as the GT deviates from the reference model. This suggests that the morphing trajectory between the open and closed forms of AK cannot be fully captured with the first normal mode of open AK. On the other hand, we see a drastic improvement for $d=16$, which suggests not only that the morphing trajectory is mostly captured with the first 16 modes but also that their amplitudes were correctly estimated by the encoder.
To quantitatively measure the accuracy of the prediction, we selected 3 predicted atomic models corresponding to the 3 local maxima in the predicted \textit{discontinuous} distribution for $d=16$ and compared them to their GT counterpart by measuring the root-mean-square deviation (RMSD) for each pair. As shown in Fig.\ref{fig:architecture}, the RMSD increases as the maxima recedes from the reference atomic model but stays below 1.5 \r{A}. For reference, the RMSD between the open and closed form is approximately 7 \r{A}.

\section{Discussion}
\label{sec:outro}
Generative modeling for heterogeneous reconstruction using atomistic representations is a nascent field in cryo-EM~\cite{Donnat2022}. It is a challenging task due to the complex nature of the conformational landscape of biomolecules, and the difficulty in estimating both this landscape and the nuisance variables of the imaging process, such as the pose of the molecule in each image or the $\ctf$ parameters. In this work, we have proposed a framework that tackles the first challenge through a simplified ENM representation of the molecule which enables a drastic reduction in dimensionality of the problem through NMA and thus makes it tractable. We have illustrated the method on synthetic datasets. 
The main drawbacks of this approach stem from its strength, namely that it relies on a reference atomic model and that the quality of the deformed models will deteriorate as they are pushed further away from the reference model \textemdash a known limitation of ENMs and NMA. These limitations are mitigated by the fact that the Protein Data Bank~\cite{berman2000protein} and the Alphafold Data Base ~\cite{varadi2022alphafold} provide relevant starting models.
While our proposed method does not estimate the orientation of the molecule, it does estimate the translational component of the pose in each image.
Future work includes demonstrating its capability on experimental datasets and tackling the second challenge with joint estimation of the molecule's conformation and pose. 

\begin{ack} We thank Mike Dunne for supporting this project. This work was supported by the U.S. Department of Energy, under DOE Contract No. DE-AC02-76SF00515, the LCLS seed grant "From Atomic Models to Noisy Images and Back Again: End-to-end Differentiable SPI simulators for CryoEM and XFELs" (PI: GW), the SLAC LDRD project "AtomicSPI: Learning atomic scale biomolecular dynamics from single-particle imaging data" (PI: FP, YN). N.M. acknowledges support from the National Institutes of Health (NIH), grant No. 1R01GM144965-02. We acknowledge the use of the computational resources at the SLAC Shared Scientific Data Facility (SDF).
\end{ack}

{\small
\bibliographystyle{plain}
\bibliography{egbib}
}

\end{document}